# Epitaxial magnetic perovskite nanostructures[**]

By *Dmitry Ruzmetov, Yongho Seo, Land J. Belenky, D. M. Kim, Xianglin Ke, Haiping Sun, Venkat Chandrasekhar*[*]*, Chang-Beom Eom, Mark S. Rzchowski,* and *Xiaoqing Pan*

Fabrication of heterostructures with atomically sharp interfaces and layer thicknesses of a few unit cells is necessary when one shrinks the size of spintronic heterostructures to the nano-scale, the scale relevant for future electronics applications. Magnetic oxide perovskites are particularly suitable for this application due to the wide range of electronic and magnetic properties available within the same crystal structure. Epitaxial perovskite multilayer heterostructures can also be grown with atomically sharp interfaces and nanometer-scale layer thicknesses.[1-3] However, the fabrication of structures from epitaxial perovskites with lateral dimensions less than a few hundred

[*] Prof. V. Chandrasekhar, Dr. D. Ruzmetov, Dr. Y. Seo
Department of Physics & Astronomy
Northwestern University
Evanston, IL 60208 (USA)
E-mail: v-chandrasekhar@northwestern.edu

L. J. Belenky, Dr. D.M. Kim, Prof. C.-B. Eom
Department of Materials Science and Engineering
University of Wisconsin-Madison
Madison, WI  53706 (USA)

X. Ke, Prof. M. S. Rzchowski
Physics Department
University of Wisconsin-Madison
Madison, WI 53706 (USA)

H. Sun, Prof. X. Pan
Department of Materials Science and Engineering
University of Michigan
Ann Arbor, MI 48109 (USA)

[**] This work is supported by NSF-ECS 0210449.

nanometers has proved challenging.[4-6] We present here a technique for the fabrication of sub-100 nm diameter magnetic perovskite nanodots that maintain their crystallinity, epitaxial structure and ferromagnetic properties after the fabrication process. Since magnetic perovskites can also be combined epitaxially with ferroelectrics and high-temperature superconductors,[7,3] this technique opens up the possibility of fabricating novel nanoscale multifunctional heterostructures.

Magnetic oxide perovskites are well known due to the phenomenon of colossal magnetoresistance (CMR) found in these materials.[8] The 100% spin-polarization of half-metallic magnetic perovskites[9] and the ability to epitaxially incorporate them into single crystal all-oxide heterostructures[1,2] are important in spintronics applications.[10] Since the materials are grown at high temperature, nanostructures from them must be fabricated by post-growth etching. The problems inherent in this fabrication process include the low etch rates of the materials, damage to the crystal structure of nanoparticles from ion milling, and redeposition of amorphous material on the substrate.[5,6]

To date, ferromagnetic perovskites have been patterned down to size scales of 300 nm,[11] while non-magnetic perovskites have been patterned down to 100 nm size scales.[12,4,13,14] In this Communication we demonstrate that by using a low energy, neutralized ion beam for milling purposes and by precisely controlling the milling timing to avoid extra damage, we are able to make sub-100 nm magnetic perovskite dots with their high quality crystalline structure and ferromagnetism preserved. These are the smallest particles made from these materials. The magnetic anisotropy of the ferromagnetic nanodots is similar to that of unpatterned films,[15] in accordance with calculations that conclude that the nanodots are still well above the superparamagnetic limit at 77 K.

The materials that we patterned into nanodots are $La_{2/3}Sr_{1/3}MnO_3$ (LSMO) and $SrRuO_3$ (SRO). LSMO is the most studied CMR manganite and a promising choice for ferromagnetic layers in magnetic tunnel junctions and other spin-injection devices since it has nearly 100% spin polarization and a high Curie temperature of 350 K.[9,16] SRO is a metallic ferromagnet with a bulk Curie temperature of 160 K.[17] SRO is often the material of choice for electrodes in the perovskite heterostructures due to its relatively low resistivity, high chemical stability and its small lattice mismatch with $SrTiO_3$, a common substrate and epitaxial base of all-oxide multilayers.[18,19]

Our nanofabrication procedure consists of the following steps. In the first step, the magnetic perovskite layers of $SrRuO_3$ and $La_{2/3}Sr_{1/3}MnO_3$ are grown on lattice matched $SrTiO_3$ (STO) and $La_{0.3}Sr_{0.7}Al_{0.65}Ta_{0.35}O_3$ (LSAT) substrates respectively in a pulsed laser deposition (PLD) system using reflection high-energy electron diffraction (RHEED)[20] to monitor *in situ* single atomic layer growth.[21] In the second step the magnetic perovskite film wafer is covered with a bilayer resist in which an array of nanometer scale holes is patterned using electron-beam lithography. The total area of a patterned array is as large as 5 $mm^2$. After development, a Ti film is evaporated onto the sample. Subsequent lift-off removes the resist leaving an array of Ti nanodots on the sample. In the last step, the sample is exposed to a beam of neutralized 200 eV argon ions. The Ti nanodots act as an etch mask during this process: the magnetic perovskite is etched except where it is protected by the Ti, resulting in an array of magnetic perovskite nanodots. The thickness of the Ti film to be evaporated is determined by the relative etch rates of the Ti and the magnetic perovskite substrate; these etch rates were determined empirically. The ion milling time was adjusted to etch completely the Ti film. We also tried other materials as etch masks, including Au, C and Ni films, e-beam resist and photoresist, but Ti gave the best results in terms of differential etch rate, resolution and adhesion to the substrate.

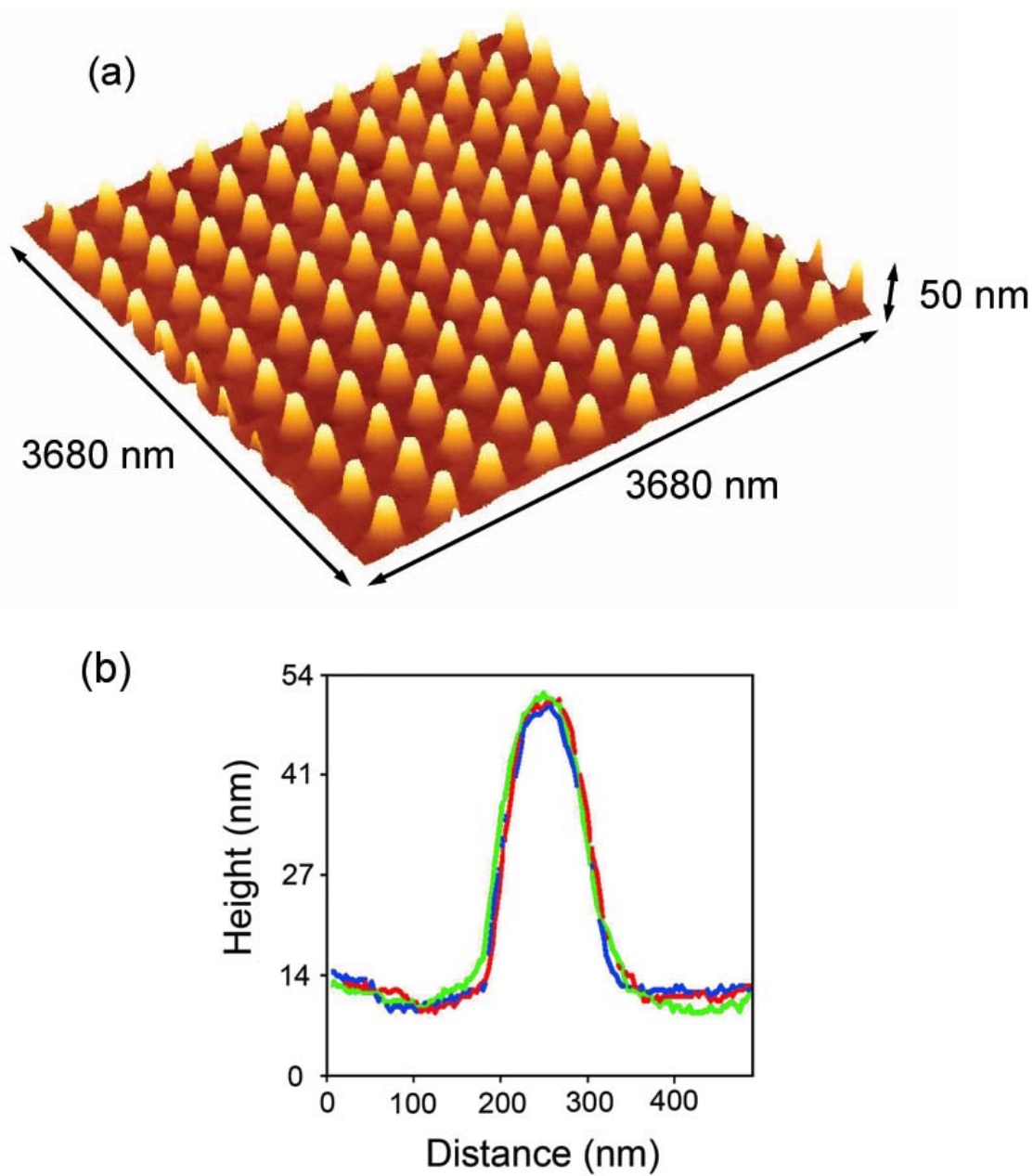

**Figure 1.** Atomic force microscope (AFM) image of a $La_{2/3}Sr_{1/3}MnO_3$ nanodot array. a) 3D AFM image of a 3680 x 3680 $nm^2$ section of the array. b) AFM profiles of three representative dots from the AFM image in (a). Dot diameters are ~100 nm, heights are 37 nm.

Figure 1(a) shows an AFM image of a section of a LSMO nanodot array. The starting substrate was a 30 nm LSMO layer on a LSAT substrate. The dots are evenly spaced, with heights of 37 nm indicating that some of the LSAT substrate is also etched during the ion milling process. Figure 1(b) shows line profiles of three randomly chosen dots in the image of Fig. 1(a), demonstrating the remarkable uniformity of the dots in the array. From these profiles, the dot diameter at half the dot height is approximately 100 nm. While AFM gives correct estimates of the height, it will overestimate the lateral size of the dot due to convolution of the tip shape with the topographic profile. In order to obtain correct estimates of the diameter of each dot, we have performed high-resolution cross-sectional Transmission Electron Microscopy (TEM) on some of our samples. Figure 2(a) shows a cross-sectional TEM image of an array of $(110)^o$ SRO[22] nanodots on a (001) STO substrate. AFM images of this sample gave diameters at half-maximum height of 110 nm, comparable to the LSMO dot array shown in Fig. 1. Cross-sectional TEM gives a diameter that is approximately half this value: the diameter of the base of each dot is about 80 nm, with a diameter at half-maximum height of 60 nm. These are the smallest particles made from ferromagnetic perovskites to date.

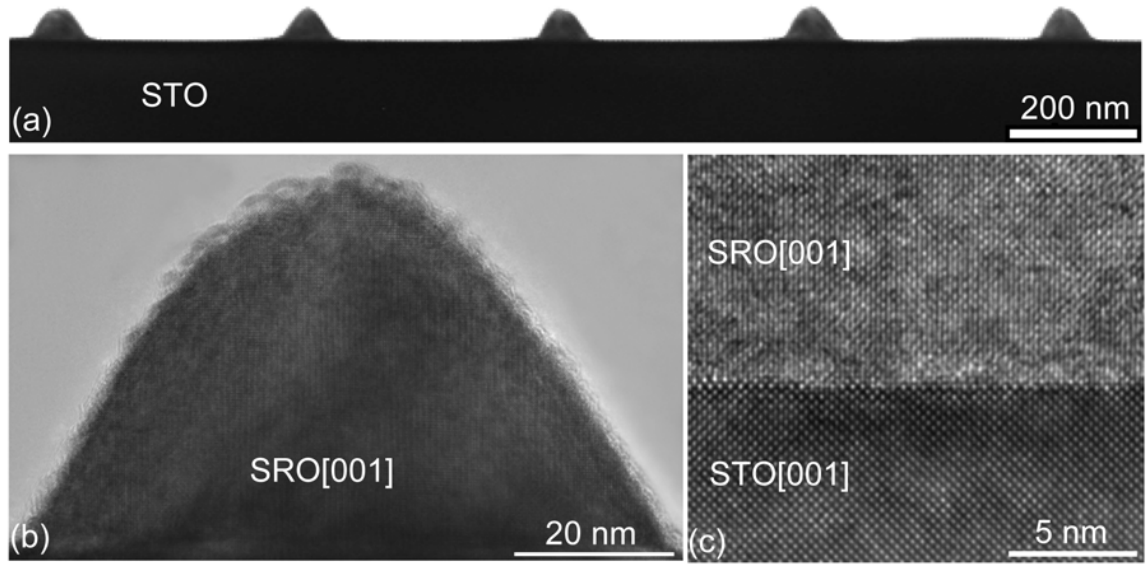

**Figure 2.** Cross-sectional TEM images of $SrRuO_3$ dots on $SrTiO_3$ substrate. a) Dot profiles. Dot material is $SrRuO_3$, lower darker region is $SrTiO_3$ substrate. b) Image of a single crystal nanodot. c) High resolution cross-TEM image of the interface between $SrRuO_3$ and $SrTiO_3$.

In addition to accurate estimates of the dot diameters, high-resolution TEM can also give information about the crystal structure of the dots. Figure 2(b) shows a TEM image of one of the dots of Fig. 2(a). X-ray energy dispersive spectroscopy (EDS) measurements confirm that the material is indeed $SrRuO_3$. There is also a 2-4 nanometer thick amorphous layer on the surface of the dot, which we believe may be due to ion damage or redeposition of the sputtered material. Figure 2(c) shows an exploded view of the interface between the SRO and the STO substrate. It can be seen that the interface is atomically sharp and the SRO lattice epitaxially matches the STO substrate; the perfect lattice planes visible in this image extend throughout this dot. Thus the TEM images demonstrate that the fabrication of the nanodot arrays preserves the crystalline nature of the substrate.

The cross-sectional TEM also gives a direct visual representation of the profile of each dot, something that is not possible to obtain with either AFM or scanning electron microscopy (SEM). If the fabrication process were ideal, one would expect cylindrical pillar-shaped dots with vertical sidewalls. The TEM images show that the profile more closely approximates a rounded pyramid. The reason for this rounded profile (which we have observed in TEMs of a few samples) is not clear. However, similar profiles have been observed in the etching of perovskite and metallic nanostructures by other groups, and have been attributed to the high energies of the ions used in the milling process.[5,4]

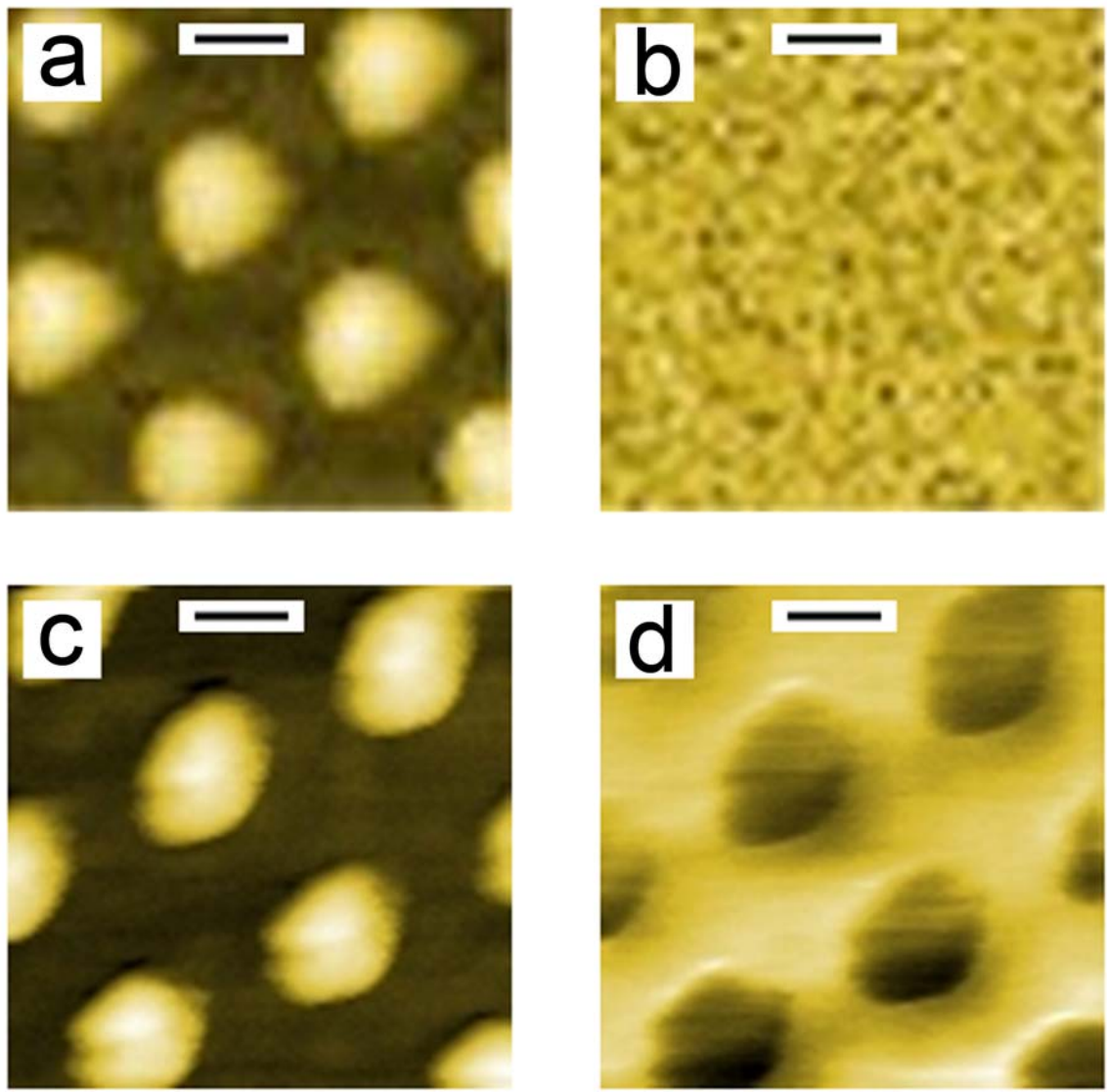

**Figure 3.** Atomic (AFM) and Magnetic (MFM) Force Microscopy images of SrRuO$_3$ nanodots at room temperature and 77K. The black scale bar corresponds to 150 nm. a) Topographic image at room temperature. b) magnetic image corresponding to (a) at room temperature showing no ordered signal since the sample is above the Curie temperature. c) and d) Topographic and magnetic images of the same sample respectively at 77 K.

In order to investigate the magnetic properties of the nanodots, we performed magnetic force microscopy (MFM) measurements on the SRO nanodot arrays. Since the Curie temperature of the two-dimensional SRO substrate is approximately 150 K, a low temperature MFM is required. The low temperature tuning fork MFM used in these measurements is described in detail elsewhere.[23] Figures 3(a) and 3(b) show topographic and MFM images of part of a SRO nanodot array taken at room temperature with this instrument. As expected, the MFM image shows no structure, since the SRO is not ferromagnetic at room temperature. Figures 3(c) and 3(d) show corresponding topographic and MFM images taken at 77 K, with the sample zero field-cooled from room temperature. The MFM image of each dot has a clear darker and a lighter region, with the darker region of all the dots being towards the bottom. This shows that each dot is magnetized, with the direction of magnetization along the vertical direction in the figure. However, the fact that all parts of each dot are darker than the surrounding substrate indicates that the magnetization does not lie in the plane of the substrate. The existence of a well-defined out-of-plane magnetization direction even in the nanometer scale dots is in agreement with the strong magnetic anisotropy observed in large SRO films, where the magnetization points at a 45° angle with the substrate.[15,24] Thus, we note that the magnetic properties of the large area magnetic perovskite films are preserved in the nanodot arrays.

Is ferromagnetism expected for the particles of this size? We calculated the anisotropy energy of a single nanodot in order to see if the superparamagnetic limit[25] was reached. Assuming an anisotropy constant $K \sim 8 \times 10^6$ erg/cc for SRO on STO at 5 K[26] as a starting point and assuming each dot to have the size shown in Fig. 2(b), we find that the magnetic anisotropy energy of a dot at 77 K is at least 4 orders of magnitude larger than the thermal energy kT. Therefore we do not expect our SRO nanodots to be near the superparamagnetic limit, as is confirmed by our MFM measurements.

In conclusion, we have demonstrated a technique for the fabrication of regular arrays of sub-100 nm magnetic perovskite structures, the smallest structures made of these materials to date. The fabrication technique described here is suitable for patterning multilayer all-oxide films into nanodevices in order to explore effects such as current driven magnetization switching,[27] tunneling magnetoresistance in magnetic tunnel junctions and current driven domain wall motion in magnetic wires,[28] as well as patterning monolayer magnetic perovskite films to study the single domain behavior of nanoparticles,[29] the superparamagnetic limit,[25] strain relaxation[30] and the consequent magnetic anisotropy change due to low dimensionality. In addition, since magnetic perovskites can be combined epitaxially into multilayer heterostructures with perovskite ferroelectrics[7] and high-temperature superconductors[3], this technique also opens up the possibility of fabricating heterostructures with new and as yet unexplored functionalities.

## *Experimental*

**Nanodot fabrication.** The process of the epitaxial growth of magnetic oxide perovskites is described elsewhere.[21] The e-beam lithography was performed with MMA(6% in Ethyl Lactate)/ 950K PMMA(4 % in Anisole) resist bilayers with MMA/PMMA thicknesses of 150 nm and 200 nm respectively, using a JEOL JSM-840 SEM at 35 kV with a beam current of around 100 pA. Patterned and developed samples were metallized in a Denton Vacuum DV-502A e-gun evaporator with a Ti deposition rate of approximately 1.2 Å/sec. The milling of the samples was done with a uniform neutralized Ar ion beam of 200 eV energy and a 16 mA beam current for various times on the order of 2 min. For nanometer scale structures under these conditions, the etch rate was 11 nm/min for the Ti films, 40 nm/min for the SRO films and 21 nm/min for the LSMO films. Thus, to define nanostructures of 70 nm height in SRO, for example, we deposited a 22 nm thick Ti film to act as an etch mask. This substrate was etched for 2 min 24 sec under the conditions described above.

**Cross-TEM measurements.** The heterostructure with $SrRuO_3$ dots was glued on to a piece of Si single crystal to protect the dot structure during the TEM specimen preparation. Thin slices of the sample were cut along the (100) plane of $SrTiO_3$ substrate. Cross-sectional specimens for transmission electron microscopy (TEM) observations were prepared by mechanical polishing and dimpling followed by ion milling. High resolution TEM studies were carried out on a JEOL3011 TEM. The composition of the nano-sized dots was analyzed by X-ray energy dispersion spectroscopy (EDS).